\documentclass[amsmath,amssymb,twocolumn,aps,pra]{revtex4-1}
\usepackage{graphicx}
\usepackage[svgnames]{xcolor}
\usepackage{bm}
\usepackage{multirow}
\usepackage{natbib}
\usepackage[colorlinks=true,linkcolor=blue,urlcolor=blue,citecolor=blue]{hyperref}

\makeatletter
\def\l@section{\@dottedtocline{0}{0.5em}{1.8em}}
\def\l@subsection{\@dottedtocline{0}{2.3em}{1.5em}}
\makeatother

\begin{document}

\markboth{Collignon et al.}{Strontium titanate}

\title{Metallicity and Superconductivity in Doped Strontium Titanate}

\author{Cl\'ement Collignon$^1$, Xiao Lin$^{1,2,3,4}$, Carl Willem Rischau$^{1,5}$, Beno\^{\i}t Fauqu\'e$^{1,6}$, and Kamran Behnia$^{1,2}$}
\affiliation{(1)Laboratoire Physique et Etude de Mat\'{e}riaux (CNRS-UPMC), ESPCI Paris, PSL Research University, 75005 Paris, France\\
(2) II. Physikalisches Institut, Universit\"{a}t zu K\"{o}ln, 50937 K\"{o}ln, Germany\\
(3) School of Science, Westlake University, 310024 Hangzhou, China\\
(4) Institute for Natural Sciences, Westlake Institute for Advanced Study, 310024 Hangzhou, China\\
(5) Department of Quantum Matter Physics, University of Geneva, 1205 Geneva, Switzerland\\
(6) JEIP, USR 3573 CNRS, Coll\`{e}ge de France, PSL Research University, 75005 Paris, France\\
}

\begin{abstract}
Strontium titanate is a wide-gap semiconductor avoiding a ferroelectric instability thanks to quantum fluctuations. This proximity leads to strong screening of static Coulomb interaction and paves the way for the emergence of a very dilute metal with extremely mobile carriers at liquid-helium temperature. Upon warming, mobility decreases by several orders of magnitude. Yet, metallicity persists above room temperature even when the apparent mean free path falls below the electron wavelength. The superconducting instability survives at exceptionally low concentrations and beyond the boundaries of Migdal–Eliashberg approximation. An intimate connection between dilute superconductivity and aborted ferroelectricity is widely suspected. In this review, we give a brief account of ongoing research on bulk strontium titanate as an insulator, a metal, and a superconductor.
\end{abstract}
\maketitle

\tableofcontents

\section{Introduction}
Strontium titanate (STO) is an insulator with a band gap of 3.22 eV \cite{Noland:1954}. It has a perovskite structure, with each Ti atom sitting at the center of a TiO$_6$ octahedron and each Sr at the center of a SrO$_{12}$ cuboctahedron. The space is filled by tessellation of these octahedra and cuboctahedra. Since Ca and Ba sandwich Sr in the same column of the periodical table, comparing SrTiO$_{3}$ with BaTiO$_{3}$ and CaTiO$_{3}$ is illuminating. Both of the latter compounds lose their cubic structure above room temperature and in different ways. BaTiO$_{3}$ is a celebrated ferroelectric insulator, undergoing a succession of structural transitions at 403, 287 and 197 K, breaking the inversion symmetry and leading to robust ferroelectricity. CaTiO$_{3}$ becomes tetragonal (but not ferroelectric), below a temperature as high as 1400 K. In contrast to its sibling compounds, SrTiO$_{3}$ keeps its cubic structure almost intact, indicating a fragile balance extremely sensitive to small perturbations.

In 1964, superconductivity with a modest critical temperature of 0.3 K was discovered in lightly-doped strontium titanate \cite{Schooley1964}. In hindsight, this discovery was a remarkable moment in the eventful history of superconductivity. This compound was one of the first 'semiconducting' superconductors \cite{Hulm}. We know now \cite{Bustarret:2015} that superconductivity is the fate of most semiconductors including archetypal silicon, diamond or germanium. It was also the first oxide perovskite superconductor, preceding the discovery of high-temperature superconductivity in cuprates by more than two decades \cite{Bednorz:1988}. Tunneling experiments in the 1980s \cite{Binnig} found two distinct gaps in this system, predating the discovery of the two-band superconductivity in MgB$_2$ in the early years of the present century \cite{MgB2}. Finally, it was found quite early \cite{Schooley1965} that the critical temperature of STO is a non-monotonic function of carrier concentration. The temperature rises first, peaks afterwards and then dives towards zero, restricting therefore, the superconducting state to a 'dome'. Since then, many other superconducting domes have been found in very different families of materials. Nowadays, the restriction of superconductivity to a finite window close to another electronic order is considered as a tantalizing clue to the origin of pairing, which is presumably boosted by the quantum critical fluctuations of its competing neighbor \cite{Laughlin:2001}.

K. A. M\"uller's 1988 Nobel lecture passionately highlighted the similarities between the two perovskite superconductors \cite{Bednorz:1988}. Nevertheless, research on superconductivity in strontium titanate faded away from the center stage of condensed-matter research in the final decades of the last century. The fundamental issue of superconductivity in doped semiconductors, first formulated by Gurevich and collaborators \cite{Gurevich1962}, remained a marginal question. In retrospect, this is not very surprising. The low critical temperature left little room for applications. As for fundamental interrogations, a theory invoking multivalley electron-phonon coupling successfully reproduced the magnitude of critical temperature and its variation with carrier concentration \cite{Koonce1967}. However, this theory was not based on the correct band structure with multiple valleys. Later band calculations identified the location of the Fermi surface at the center of the Brillouin zone \cite{MattheissI}. This was confirmed much later by experimental fermiology \cite{Uwe,Lin2013,Allen,Lin2014}. STO is now firmly established as not being a multivalley electronic system.

Even in the 1980s, not everybody admitted that the issue of superconductivity in strontium titanate was settled. In particular, Takada \cite{Takada:1980} pointed to the fact that because of the small Fermi temperature compared to the Debye temperature, a straightforward application of the BCS theory is problematic. Using a frame elaborated by Kirzhnits, Maksimov and Khomskii \cite{Kirzhnits1973}, he calculated a critical temperature with a dome-like structure as a function of carrier density. Appel \cite{Appel} wondered if the soft mode associated with the structural transition at 105 K plays a role in pairing. With a few exceptions \cite{Binnig}, however, experimental research on superconducting strontium titanate was no longer alive.

In 2004, Ohtomo \& Hwang discovered a metallic two-dimensional electron gas at the interface between SrTiO$_3$ and LaAlO$_3$ \cite{Ohtomo:2004}. This led to the discovery of superconductivity of this metallic interface three years later \cite{Reyren:2007} and a new field of research. Soon it became clear that the heterointerface between SrTiO$_{3}$ and numerous insulators (including vacuum \cite{Santander:2007}) is metallic and whenever this two-dimensional metal can be cooled below 0.3 K, it becomes a superconductor \cite{Biscaras:2010}. The intense research activity on interface superconductivity generated renewed interest in the origin of its bulk parent.

Five decades after its discovery, strontium titanate remains by far the most dilute superconductor among doped semiconductors \cite{Bustarret:2015}. The persistence of superconductivity in such a context raises several questions: Can one apply the BCS theory when the hierarchy between Fermi and Debye energies is reversed? How can we conceive phonon-mediated superconductivity in the absence of retardation? How can we reconcile the Macmillan formula with a finite critical temperature of 0.1 K when the density of states corresponds to one electron per 50,000 formula units (f.u.)? It is an unavoidable temptation to speculate about a connection between dilute superconductivity and aborted ferroelectricity \cite{Muller:1979}. According to recent experiments \cite{Lin2014PRB,Lin2015PRB}, the superconducting order parameter has a purely s-wave symmetry at least at optimal doping. Several authors have suggested quantum critical scenarios \cite{Rowley:2014,Edge:2015} based on the idea of ferroelectric quantum criticality \cite{Roussev:2003}. Some experimental support for this has recently emerged \cite{Stucky:2016,Rischau:2017}.

Electronic transport after chemical substitution in the parent insulator generates another set of questions. Introducing mobile electrons to this insulator is remarkably easy, in contrast to hole-like carriers. This dissymmetry is presumably due to the indirect gap, which puts the $n$-doped and $p$-doped constant-energy surfaces in different locations in the Brillouin zone. The precocious metallicity of the $n$-doped semiconductor is understandable in the light of the Mott criterion for metal-insulator transitions \cite{Edwards:1978}, which states that the critical doping for metal-insulator transition, $n_{c}$ is set by the effective Bohr radius, $a^{*}_{B}$, namely: $n_{c} \, a^{*}_{B} \cong 0.25$. Here, $a^{*}_{B}$ is unusually long because of the large static electric permittivity \cite{Muller:1979}. A long $a^{*}_{B}$ can also explain why the low-temperature electron mobility is large \cite{Behnia:2015}, even in the absence of controlled homogeneous doping. The temperature dependence of resistivity of this dilute metal was the subject of forgotten debates in the 1960s \cite{Frederikse:1967,Wemple:1969}. At low temperature, resistivity is quadratic in temperature \cite{MarelI,Lin:2015} as is expected in a Fermi liquid. But does this temperature dependence simply reflect the size of the phase space for electron-electron scattering? At high-temperatures the exponent becomes larger than two and the behavior does not match the Bloch-Gr\"uneisen picture of electron-phonon scattering. Recently, it has been found that injecting the well-established low-temperature effective mass to the Drude expression for conductivity at room temperature implies an unrealistically short carrier mean-free-path \cite{Lin:2017}. Is the temperature dependence of resistivity simply set by the change in the scattering rate of carriers? If so, then this is another bad metal, which keeps its metallicity beyond the Mott-Ioffe-Regel limit. Or is the effective mass of the polaronic carriers changing and becoming heavier with warming? There are no consensual answers to these questions yet.

This review tries to give a brief account of the accumulated knowledge on this solid as an insulator, a metal and a superconductor. The focus is on the bulk solid. The interested reader can find many reviews on the superconducting properties of hetero-structures \cite{Hwang:2012,Zubko:2011}. A more detailed review of phenomena associated with ferroelectric quantum criticality can be found in \cite{Chandra:2017}.

\section{The Crystal Lattice of SrTiO$_{3}$}
\begin{figure*}
\centering
\includegraphics[width=0.9\textwidth]{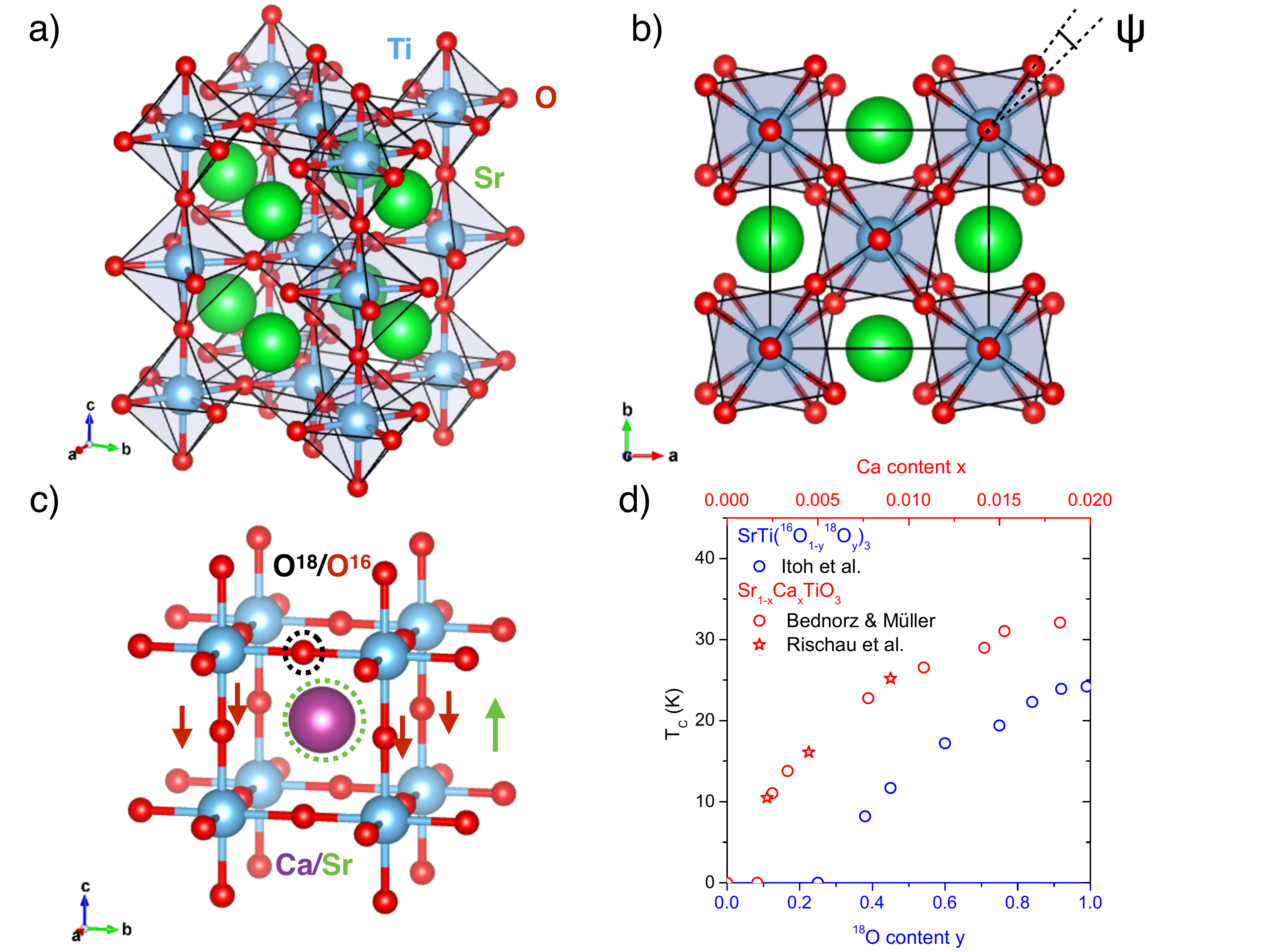}
\flushleft
\caption{\textbf{Structure and instabilities of SrTiO$_3$ a)} Lattice structure in the tetragonal phase, the titanium atoms are in the center of oxygen polyhedras. \textbf{b)} View along the c-axis of the antiferrodistortive phase. The unit cell is
shown in black. \textbf{c)} Substituting $^{16}$O with $^{18}$O or Sr with Ca favors the ferroelectric instability, Ca atoms are smaller than Sr and $^{18}$O heavier than $^{16}$O. \textbf{d)} Curie temperature, $T_{Curie}$, determined from dielectric measurements in SrTi${({}^{16} \textnormal{O}_{1-y}{}^{18} \textnormal{O}_{y})}_3$ \cite{Itoh:2003} and Sr$_{1-x}$Ca$_{x}$TiO$_{3}$ \cite{Bednorz:1984, Rischau:2017} as a function of Ca and $^{18}$O content, respectively. }
\label{Fig1}
\end{figure*}

At room temperature, SrTiO$_{3}$ has a cubic structure (Figure \ref{Fig1}a), with $Pm\overline{3}m$ symmetry and a lattice parameter of $a = 3.9053$ \AA. It undergoes a structural transition at $T_{AFD}=105$ K and becomes tetragonal. This transition, called antiferrodistortive (AFD) because of the antiphase rotation of the TiO$_6$ octahedra (Figure \ref{Fig1}b), has played an important role in the history of research on structural phase transitions. It was the first case of a zone-boundary soft mode \cite{MullerBook:1981}. SrTiO$_{3}$ is subject to another lattice instability associated with the softening of a zone-center phonon mode, potentially leading to a ferroelectric distortion. The softening of this mode remains incomplete even at zero temperature. This soft mode is the origin of the quantum paraelectric phenomenon.

Density functional theory has confirmed the instability of the cubic structure. The experimental lattice parameter leads to imaginary phonon frequencies both at the boundary (R-point) and the center ($\Gamma$-point) of the Brillouin zone \cite{Aschauer:2014}. Recently, two theoretical groups have employed alternative techniques, based on anharmonic lattice dynamics \cite{Feng:2015} or microscopic anharmonic force constants \cite{Tadano:2015} and obtained a phonon spectrum free of imaginary frequencies.

\subsection{The Antiferrodistortive (AFD) Transition}
Early X-ray measurements quantified the slight tetragonal distortion $c/a$= 1.0005 caused by the the AFD transition without any detectable change in the unit-cell volume \cite{Lytle:1964}. Based on Electron Spin Resonance measurements, Unoki \& Sakudo suggested that the unit cell in the tetragonal phase is $\sqrt{2}a\times\sqrt{2}a \times 2c$ (shown in Figure \ref{Fig1}b). This implies that each pair of adjacent TiO$_6$ octahedra rotates clock-wise and anti-clockwise around the c-axis \cite{Unoki:1967}. This scenario was confirmed by Raman \cite{Fleury:1968} and neutron \cite{Shirane:1969} scattering measurements,which also showed that the [111] zone-boundary phonon, located at the R-point softens. In the tetragonal phase, the triply-degenerate soft mode splits into a singlet for rotations around the $c$-axis and a doublet for rotations around the $a$ and $b$ axes. The octahedra rotation angle is as small as $\phi=1.4^{\circ}$ \cite{Unoki:1967, Shirane:1969} at $T=70$ K. A review of research on the AFD transition in the three decades preceding 1996 can be found in ref. \cite{Cowley:1996}. A detailed study of the specific heat near the critical temperature, found that this phase transition is second-order, mean-field and close to tri-criticality \cite{Salje:1998}.

The octahedra can rotate around each of the three perpendicular cubic axes. Therefore, in absence of controlled strain, three types of tetragonal domains along three different $c$-axes emerge. Single-domain samples have been obtained by manipulating the shape of the crystal or applying directional stress \cite{Muller:1970}. Detailed studies of the lattice constant of single-domain samples by high-resolution (better than 1 ppm) X-ray diffractometry found that the TiO$_6$ polyhedron does not remain a perfect octahedron, but slightly elongates along its c-axis \cite{Sato:1985}.

The AFD transition is extremely sensitive to the presence of a tiny amount of extrinsic atoms. Introducing oxygen vacancies reduces $T_{AFD}$ \cite{Bauerle:1978,Tao:2016}. However, substituting Ti with Nb shifts it upward. Substitution of Sr with La or Ca is beneficial for $T_{AFD} $ \cite{Tao:2016,Lima:2015}, but it is detrimental with Ba and Pb \cite{Smirnova:2005}. McCalla et al. \cite{McCalla:2016} have recently examined in detail the complex response of $T_{AFD}$ to substitution and proposed a unified picture to explain it that is based on a combination of impurity-induced change in the Goldschmidt tolerance factor and the ionic valence mismatch between the host and the impurity.

\subsection{A Quantum Paraelectric}
\label{QP}
SrTiO$_3$ is subject to a second lattice instability associated with a soft mode at the $\Gamma$-point (the $\Gamma_{15}$ mode), which could lead to a ferroelectric instability. The softening of the $\Gamma_{15}$ mode has been observed by Raman \cite{Fleury:1968b} and neutron \cite{Yamada:1969} scattering. In contrast to the AFD transition (where the softening of the R-mode is complete at $T_{AFD}$), the frequency of the transverse optical mode never reaches zero frequency. The system is therefore very close to a ferroelectric instability without effectively ordering.

This \textit{incipient ferroelectricity} \cite{Uwe:1976} shows itself in its large dielectric constant. As the system is cooled from room temperature to liquid He, the dielectric constant rises from 300 to 20,000 times vacuum permittivity \cite{Weaver:1959}. Below about 40 K, it deviates from a Curie law and saturates \cite{Muller:1979}. This is a quantum paraelectric state caused by the stabilization of large zero-point quantum fluctuations preventing ferroelectricity.

In a polar crystal, there is a theoretical link between the ratio of static, $\epsilon_{0}$, to high-frequency, $\epsilon_{\infty}$, electric permittivity and the ratio of longitudinal, $\omega_{LO}$, to transverse, $\omega_{TO}$, phonon frequencies known as the Lyddane-Sachs-Teller expression \cite{Cochran:1962}:

\begin{equation}\label{6}
\frac{\epsilon_{0}}{\epsilon_{\infty}}={\left(\frac{\omega_{LO}}{\omega_{TO}}\right)}^2
\end{equation}

The scaling between static electric permittivity, $\epsilon_{0}$, and the frequency of the transverse optical (TO) soft mode \cite{Fleury:1968,Yamada:1969}, $\omega_{TO}$ has been experimentally verified. Yamada \& Shirane \cite{Yamada:1969} showed that $\omega_{TO}=194.4\sqrt{\epsilon^{-1}}$ within experimental resolution, where $\omega_{TO}$ is expressed in meV. Given that the highest-frequency Longitudinal Optical (LO) phonon corresponds to 100 meV, this would imply an $\epsilon_{\infty}\simeq 4 $. Thus, the large static screening disappears at high frequencies.

\subsection{Multiple Roads Towards Ferroelectricity}
As discussed above, SrTiO$_3$ is a quantum paraelectric. Its paraelectric state shows the soft-mode behavior of a displacive ferroelectric. The potential energy of the polar soft mode as a function of the relative displacement of anions (O) and cations (Sr, Ti) exhibits a double well shape with two minima corresponding to ferroelectric ground states with opposite polarizations. In pristine SrTiO$_3$ zero-point fluctuations between those two states impede the emergence of a long-range ferroelectric order. This order can be stabilized by chemical substitution (of a small fraction of Sr atoms with Ca \cite{Bednorz:1984} or Ba \cite{Lemanov:1996}), by isotopic substitution (of ${}^{16}$O with ${}^{18}$O \cite{Itoh:1999}) or by application of stress \cite{Uwe:1976} or an electric field \cite{Hemberger:1995}.

Sr$_{1-x}$Ca$_{x}$TiO$_{3}$ (CaSTO) becomes ferroelectric for Ca concentrations of $0.002<x<0.02$ and it's Curie temperature $T_C$ increases with increasing $x$ as shown in Figure \ref{Fig1}d. It should be noted that the complete phase diagram of CaSTO is complex with multiple structurally-distinct phases \cite{Carpenter:2006}. The ferroelectric phase is restricted to a very narrow window with a calcium content lower than two percent. Dielectric and linear birefringence measurements on CaSTO showed that CaSTO is an XY-ferroelectric with the polarization building up along the $[110]$ and $[1\bar{1}0]$ directions \cite{Bednorz:1984,Kleemann:1997}. An activation of TO phonon modes below $T_C$ has been observed with Raman spectroscopy \cite{Kleemann:1997}. In SrTi${({}^{16} \textnormal{O}_{1-y}{}^{18} \textnormal{O}_{y})}_3$ (STO18) ferroelectricity appears above an ${}^{18}$O substitution level of around 35\%, evidenced by the appearance of a peak in the dielectric constant \cite{Itoh:1999,Itoh:2003} and a hardening of the TO$_1$ soft-mode seen in Raman measurements below $T_C$ \cite{Shigenari:2006}. Figure \ref{Fig1}d compares $T_C$ measured in both CaSTO and STO18.

The emergence of ferroelectricity in CaSTO and STO18 has been described using two different pictures, since the transition exhibits components of both displacive and order-disorder type ferroelectrics. In CaSTO, the substituted Ca atoms have a slightly smaller ionic radius (0.99 {\AA}) than Sr (1.12 {\AA}) (see Figure \ref{Fig1}c). In the first (order-disorder or percolation) picture, Ca atoms take off-center positions and form polarized dipole clusters that grow in size with decreasing temperature. The percolation of these clusters at $T_C$ leads to a long-range ferroelectric order \cite{Bianchi:1994}. In the second picture, Ca substitution damps the quantum fluctuations that prevent ferroelectricity in pure STO. A transverse Ising model has been used to describe the competition between the zero-point quantum fluctuations and the ferroelectric order \cite{Wang:1998}. In this model, the dipolar interaction between Ca sites enhances the cooperative exchange coupling between dipoles and suppresses tunneling between the possible ferroelectric ground states. In STO18, substitution of ${}^{16}$O with larger mass ${}^{18}$O atoms (Figure \ref{Fig1}c) lowers the zero point energy in the ferroelectric double well potential and tunneling between the two ground states is reduced, thus enabling a condensation of the polar soft mode \cite{Edge:2015}. In addition to this soft mode behavior, an order-disorder component of the transition was reported based on the observation of polar clusters in NMR measurements \cite{Blinc:2005}.

\subsection{Phonon Hydrodynamics}
Two rare transport phenomena, second sound and Poiseuille flow, manifestations of phonon hydrodynamics \cite{Beck:1974} were observed in this insulator and traced back to the presence of soft modes in the phonon spectrum. Second sound is a wave-like propagation of temperature or entropy. Studies of Brillouin scattering \cite{Hehlen:1995} and low-frequency light scattering \cite{Koreeda:2007} have concluded that it occurs in this solid. Poiseuille flow is invoked when the temperature dependence of thermal diffusivity is set by the variable viscosity of the phonon fluid. It has been detected by a study of thermal conductivity \cite{Martelli:2018}. Both these signatures of phonon hydrodynamics require a combination of frequent Normal scattering and rare resistive (that is, Umklapp and impurity) collisions. Strong coupling between soft phonons and acoustic phonons \cite{Bussmann:1997} can enhance Normal scattering. This is invoked to explain the observation of phonon hydrodynamic in this solid \cite{Koreeda:2007,Martelli:2018}.

\section{Metallicity of Doped SrTiO$_{3}$}
\subsection{Charge Transport: Historical Background}
The first report on electronic transport in doped strontium titanate \cite{Frederikse:1964} and the latest discussion of it \cite{Lin:2017} are separated by fifty-three years. During this time, numerous papers treated the subject \cite{Tufte:1967, Frederikse:1967, Wemple:1969, Baratoff:1981, MarelI, Verma:2014, Himmetoglu:2014, Lin:2015, Mikheev:2015}. Surprisingly, however, there is no consensual picture of the way electrons travel in this dilute metal.

Early studies found that the electric resistivity changes by orders of magnitude when one cools the solid from 300 to 4 K \cite{Frederikse:1964, Tufte:1967, Frederikse:1967, Wemple:1969}. This was very different from the temperature dependence of resistivity in metallic silicon \cite{Yamanouchi:1967}. Cooling doped silicon at a carrier concentration of $6 \times 10^{18}$ cm$^{-3}$, leads to a twofold increase in mobility \cite{Yamanouchi:1967}. In doped strontium titanate at the same carrier concentration, the mobility enhancement is thousandfold \cite{Lin:2017}. A room temperature to residual-resistivity-ratio of 1000 is not surprising per se. It has been observed in numerous metals. Here, however, a thousandfold change in the scattering time (or equivalently in the mean-free-path) given our knowledge of the zero-temperature properties of the system, generates a paradox.

\subsection{Band Structure and Experimental Fermiology}
\label{Sec3.2}
Experimental fermiology \cite{Lin2013,Lin2014,Allen} agrees with the expectations of \textit{ab initio} band calculations \cite{MarelI,Tao:2016}. According to theory \cite{MarelI}, stoichiometric SrTiO$_3$ has an indirect gap at the $\Gamma$-point. The filled bands are associated with oxygen $2p$ orbitals and three empty bands originate from Ti $3d$ orbitals. In the cubic state and in the absence of spin-orbit coupling the latter three bands are degenerate. The tetragonal crystal field and the finite spin-orbit coupling lift the degeneracy and create three distinct bands all centered at the $\Gamma$-point \cite{MarelI}. The lower band has a distinct non-parabolic dispersion. It starts as a light band and becomes heavier when $k>0.1/a$ as a consequence of band anti-crossing. The non-parabolic dispersion of the lower band is also captured by tight-binding models.

Like what was reported in other dilute metals \cite{Behnia:2007,Zhu:2010,Fauque:2013}, measurements of the Nernst effect \cite{Lin2013},found giant quantum oscillations. The oscillating component is much larger than the background. Extensive Shubnikov-de Haas studies \cite{Lin2014} detailed the variation of the multiple frequencies with carrier concentration. A single frequency was found in the dilute limit \cite{Lin2013}. Assuming a single slightly anisotropic pocket, this frequency yielded a carrier concentration matching the Hall number. This establishes that the Fermi surface is indeed located at the $\Gamma$-point. As the carrier concentration increases, this Fermi surface grows in size and above a threshold concentration, $n_{c1}$, close to what the theory had suggested \cite{MarelI}, a second small frequency, pointing to the emergence of an additional concentric pocket, is detected \cite{Lin2014}. The outer Fermi surface becomes heavier when $n>n_{c1}$, confirming another expectation of \textit{ab initio} theory. An angle-dependent study of quantum oscillations in La-doped SrTiO$_3$ thin films confirmed the expected anisotropy of the lower band at low carrier concentration \cite{Allen}. The success of the rigid-band picture using virtual crystal approximation in describing non-stoichiometric strontium titanate is remarkable.

The experimentally-resolved mass at low concentration (1.8$m_e$) \cite{Lin2013,Lin2014} is 2.5 times the expected band mass (0.7$m_e$) \cite{MarelI}, suggesting a finite but modest mass renormalization. A mass renormalization of similar magnitude was obtained by infrared conductivity \cite{Mechelen:2008}. With increasing carrier concentration , the cyclotron mass of the outer Fermi surface increases and becomes as heavy as 3.5$m_e$ \cite{Lin2014}, which is what is expected by the theoretical non-parabolic dispersion of the lower band \cite{MarelI}.

A further check of the overall consistency is provided by specific heat data, which resolved an electronic specific heat of $\gamma = 1.55$ mJ/K$^2$cm$^{-3}$ \cite{Lin2014PRB} at a carrier concentration of $n=2.6\times 10^{20}$ cm$^{-3}$. Now, quantum oscillations quantify how carriers are distributed among three pockets of unequal size. Using this information and the three different cyclotron masses, one finds $\gamma = 1.53$ mJ/K$^2$cm$^{-3}$ in excellent agreement with the measured value. This leaves little doubt on the accuracy of the band picture in describing the electronic properties at low temperature.

\subsection{Bohr Radius, Metal-Insulator Transition and High Mobility}
Quantum oscillations can be observed in this doped semiconductor in presence of a moderate magnetic field. Since this is not the case for metallic silicon, one may wonder why. It is because the parent insulator of dilute metallic strontium titanate is a quantum paralectric. The effective Bohr radius ($a^*_B=4\pi \epsilon_{0}\hbar^{2}/m^{*}e^{2}$) depends on the static permittivity, $\epsilon_0$, and the effective mass, $m^{*}$. Because of the large magnitude of $\epsilon_0$, $a^*_B$ becomes as long as 600 nm in metallic STO \cite{Behnia:2015}. According to the Mott criterion for a metal insulator transition, metallicity \cite{Edwards:1978} should emerge at a carrier concentration as low as $n_c\simeq 10^{11}$ cm$^{-3}$. Unavoidable extrinsic impurities, impede a controlled study of a metal-insulator transition at such carrier concentrations \cite{Spinelli:2010}. However, in such a context, observing a sharp Fermi surface down to 10$^{17}$ \cite{Lin2013,Allen,Lin2013} or even 10$^{16}$ cm$^{-3}$ \cite{Bhattacharya} becomes unsurprising.

How can a shallow Fermi sea (with a depth of 1 meV on top of a band gap of 3 eV) survive without disintegrating into a collection of puddles in absence of homogeneous distribution of oxygen vacancies? The long effective Bohr radius plays a crucial role by pushing up the Thomas-Fermi screening length in the dilute metal. This protects the percolated Fermi sea even when its depth becomes more than three orders of magnitude smaller than the gap and keeps the mobility of the carriers high, because any local departure from lattice perfection is averaged over long distances. The doping dependence of mobility, which follows $n^{-\alpha}$ with $\alpha$ close to unity, can be explained with a set of most unsophisticated assumptions, leading to an expression for the intrinsic mobility of a metallic semiconductor with a random distribution of dopants \cite{Behnia:2015}.

\subsection{T-Square Resistivity}
\begin{figure*}
\centering
\includegraphics[width=0.95\textwidth]{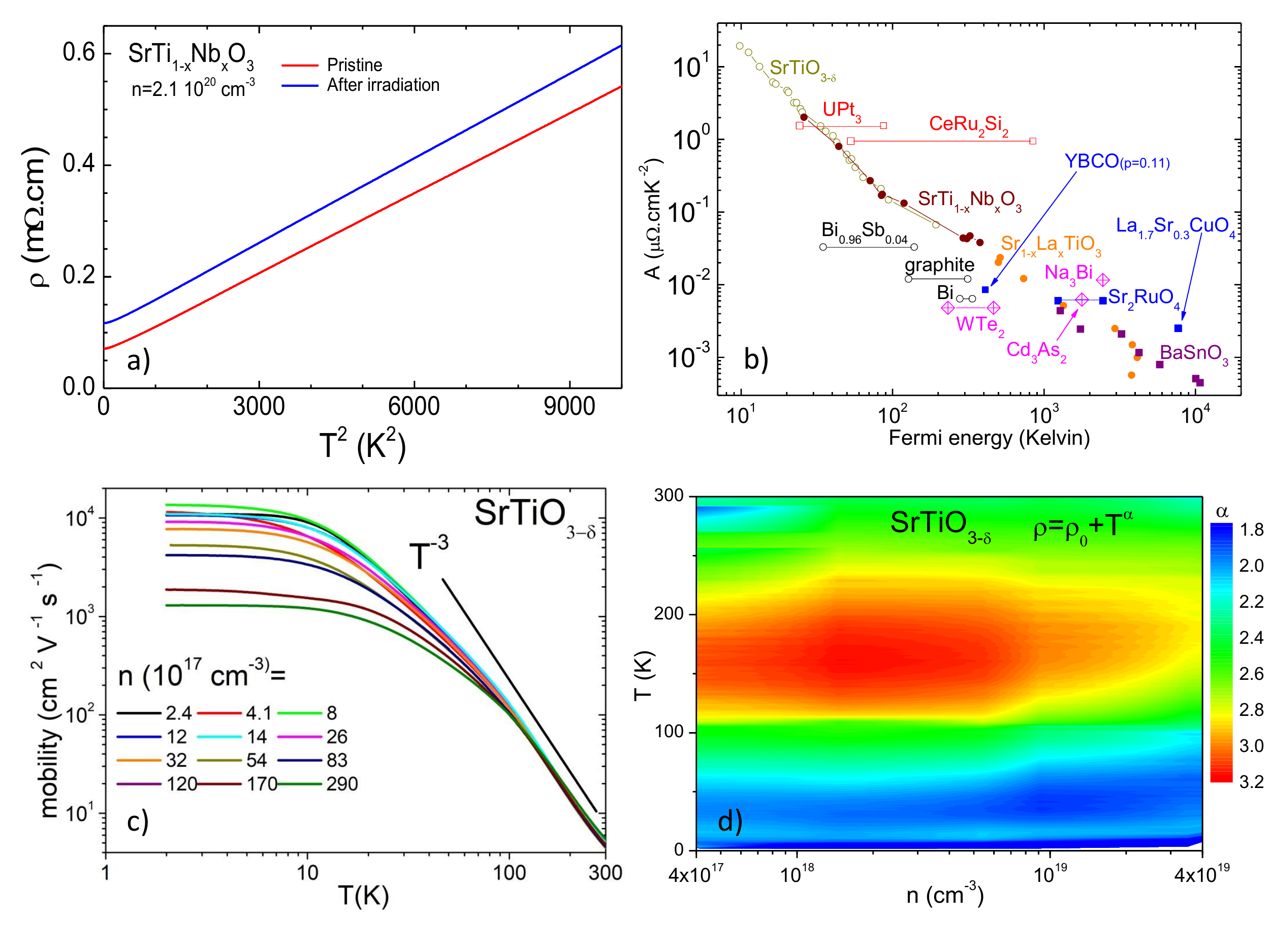}
\flushleft
\caption{\textbf{Resistivity of doped STO} \textbf{a)} Temperature dependence of the resistivity of $n$-doped SrTiO$_3$ following $\rho=\rho_0+ AT^2$. The slope, i.e., the prefactor $A$ shows little change after irradiation by high-energy electrons introducing disorder \cite{Lin2015PRB} \textbf{b)} The prefactor $A$ of  $T$-square resistivity as a function of the Fermi energy for a number of dilute metals. $A$ scales inversely with the Fermi energy \cite{Lin:2015}; this is an extension of the Kadoaki-Woods \cite{Kadowaki} scaling in low-density metals. \textbf{c)} Mobility as a function of temperature. Note the existence of two different regimes. In the low-temperature regime, mobility depends on Fermi energy and carrier concentration, but above 100 K, mobility becomes independent of carrier concentration and follows a temperature dependence close to  $T$-cube. \textbf{d)} Color plot of the resistivity exponent $\alpha$ in the (T, n) plane. Note that $\alpha=2$ only holds at low temperatures and that $\alpha$ exceeds 3 at intermediate temperatures before decreasing again.}
\label{Fig2}
\end{figure*}

The Tokura group, studying the phase diagram of Sr$_{1-x}$La$_x$TiO$_3$, was the first to show that the low-temperature resistivity of the dilute metal follows a  $T^2$ behavior \cite{Okuda:2001}. Their study was restricted to carrier densities exceeding 0.01 e$^-$ per formula unit (f.u.). Subsequently, van der Marel et al. showed that  $T$-square resistivity persists down to carrier densities as low as $10^{-3}$e$^-$ per f.u. \cite{MarelI}. Then, Lin et al. showed that the behavior persists even when the carrier density becomes as low as $3 \times 10^{17}$ cm$^{-3}$ (corresponding to $2 \times 10^{-5}\,$e$^-$ per f.u.) \cite{Lin:2015}. Across a concentration window extending over four orders of magnitude, the low-temperature electric resistivity can be expressed as $\rho=\rho_0+ AT^2$. The residual resistivity, $\rho_0$, is enhanced when one introduces controlled disorder \cite{Lin2015PRB} by irradiating a sample, but $A$ is not (see Figure \ref{Fig2}a). It smoothly increases with decreasing carrier concentration, becoming as large as a few $\mu \Omega$cmK$^{-2}$, typical of a heavy-electron metal (see Figure \ref{Fig2}b).

One expects a  $T$-square resistivity in a Fermi liquid, because the phase space for electron-electron scattering \cite{Mott} has such a temperature dependence. This phase space is inversely proportional to the square of the Fermi energy, therefore, the smooth increase in the magnitude of the prefactor with decreasing Fermi energy (which scales with carrier concentration) is also expected (Figure \ref{Fig2}b). At a first glance, all this appears to be a simple extension of the Kadowaki-Woods scaling \cite{Kadowaki} to a low-density metal.

However, the observation raises several questions. There are two known ways for collision between electrons to generate a finite contribution to electric resistivity. The first is Umklapp scattering, during which the colliding electrons lose a unit vector of the reciprocal lattice. But in the extreme dilute limit, the Fermi surface is too small for Umklapp scattering. The second possibility is often dubbed the Baber mechanism \cite{Baber}. If there are two reservoirs of electrons, with one being more coupled to the lattice, even if the momentum is ultimately lost by electron-phonon collisions, the rate is set by a bottleneck set by the momentum exchange rate between the two reservoirs. However, as discussed above, there is only one band in the extreme dilute limit.

A second problem is the persistence of the  $T$-square behavior up to the degeneracy temperature and above. In the Fermi liquid picture, one expects a  $T$-square behavior only when the quasi-particles are well-defined and deep in the degenerate regime. Maslov \& Chubukov have recently pointed out \cite{Maslov2017} that there is an old and alternative explanation for the  $T$-square resistivity \cite{Epifanov1981} invoking the collision of electrons with two TO phonon modes. One shall not forget, however, that the  $T$-square resistivity is the asymptotic low-temperature behavior and accounts for a tiny fraction of the overall temperature-dependent resistivity. The exponent is smoothly evolving (Figure \ref{Fig2}d) and this blurs a reliable detection of the temperature above which  $T$-square resistivity stops.

\subsection{T-Cube Resistivity}

Above 100 K, the temperature dependence of resistivity is close to cubic, as reported in early studies \cite{Tufte:1967,Frederikse:1967} and confirmed by more extensive recent measurements \cite{Lin:2015,Lin:2017}. In this regime, the mobility becomes independent of carrier concentration. It is equal to $4.9\pm 0.5$ cm$^2$/Vs at 300 K (see Figure \ref{Fig2}c) and continues to decrease as $T^{-\alpha}$, with $\alpha>2$ with warming (see Figure \ref{Fig2}d). Given what we know about the low-temperature properties, this metallicity defies the Boltzmann picture. Injecting the carrier concentration and the effective mass resolved by quantum oscillations \cite{Lin2014,Allen}, the carrier mean-free-path becomes significantly shorter than the de Broglie thermal wavelength and the interatomic distance. In other words, this metallicity does not respect the boundaries set by the Mott-Ioffe-Regel limit \cite{Hussey}.

In 1981, Baratoff \& Binnig \cite{Baratoff:1981} fitted the resistivity over the entire temperature range by adding to the  $T$-square term a sum of two weighted Bose-factors representing the population of two LO phonon modes at 58 and 100 meV. In this picture, the main driving force behind the temperature dependence of resistivity is the change in the population of thermally excited LO phonons strongly scattering electrons. More than three decades later, Micheev et al. \cite{Mikheev:2015} used a similar approach. This led to an excellent fit, but raised several unanswered questions: Why does the mean-free-path continues to decrease even after falling below the carrier wavelength? What explains the persistence of the  $T$-square component up to room temperature (which is required in addition to the Bose factor to obtain a satisfying fit)? Why does metallic BaTiO$_{3-\delta}$ \cite{Kolod}, which has similar LO phonons, have a very different temperature-dependent resistivity?

Although SrTiO$_{3-\delta}$ behaves very differently from ordinary doped semiconductors, including those with a similar hard phonon spectrum, its temperature-dependent resistivity is strikingly similar to other aborted ferroelectrics such as IV-VI semiconductors \cite{Allagier:1958} and doped KTaO$_3$ \cite{Wemple1965,Sakai2009}. This observation points to the soft TO phonons as the driving force behind the temperature dependence of resistivity \cite{Lin:2017}. As early as 1966, Wemple, invoking the striking similarity in the effect of pressure on resistivity and electric permittivity in various perovskites, argued in favor of a central role played by TO phonons in charge transport \cite{Wemple1966}.

If the quasi-particles do not change their mass with temperature, then they cease to be well-defined over a time scale long enough to allow a scattering time, used in the collision-based Boltzmann picture. This is the central paradox for any collision-based picture of charge transport in this system. But what if the polarons were becoming heavier upon warming? This interesting line of thought was suggested by Eagles \cite{Eagles1996} who invoked the thermal evolution of plasma frequency as reported by Gervais \emph{et al.} \cite{Gervais1993}. Later and more extensive studies by van der Marel's group \cite{Mechelen:2008} led to a quantitatively different set of data and an attenuated temperature dependence. It is important to recall what Fredrikse \emph{et al.} noticed as early as 1964 \cite{Frederikse:1964} that small polarons are not expected to become heavier with warming.

\section{The Most Dilute Superconductor}

\begin{figure*}
\centering
\includegraphics[width=0.95\textwidth]{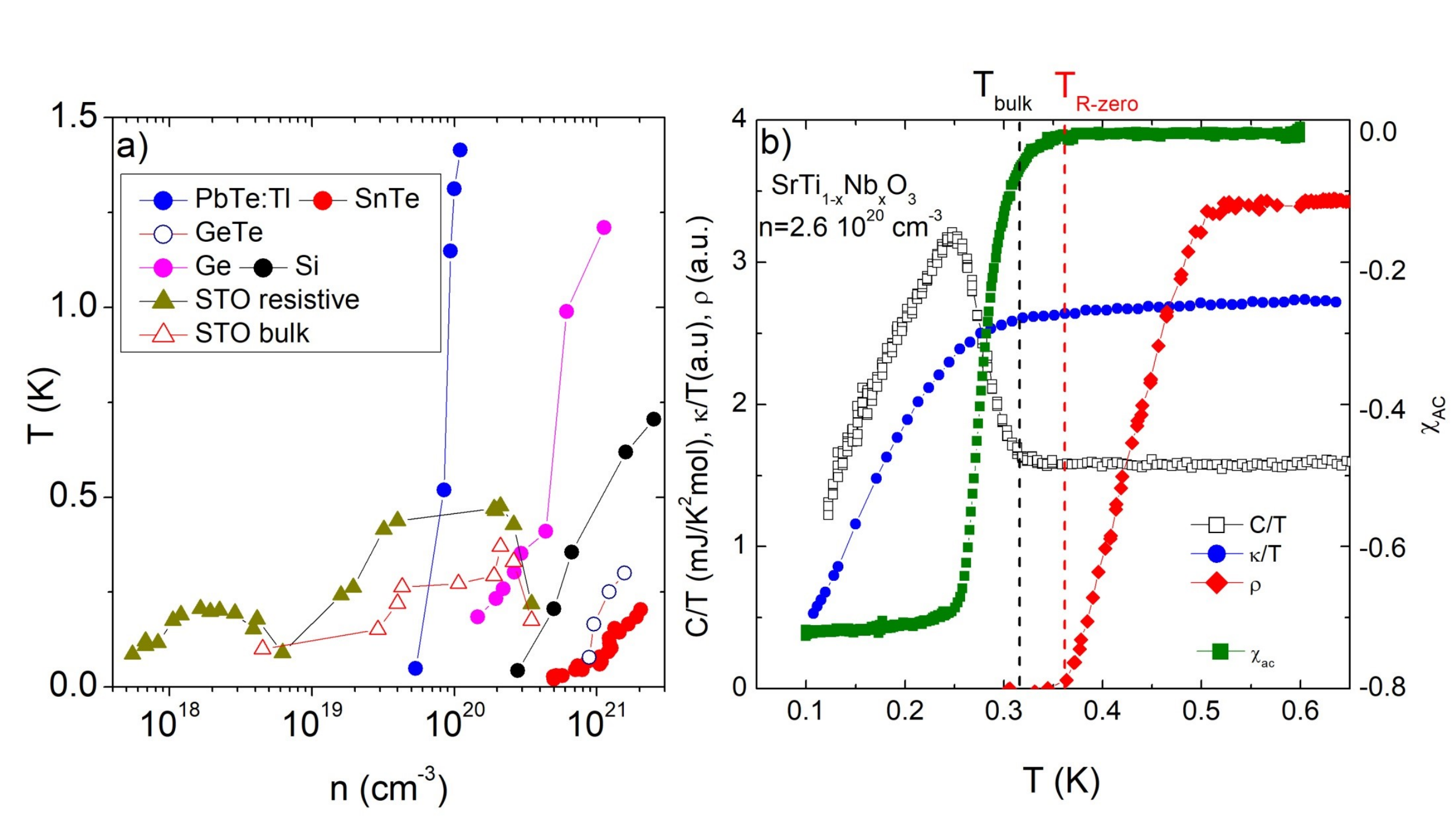}
\flushleft
\caption{\textbf{Features of superconductivity in doped STO} \textbf{a)} Superconducting phase diagram for a group of semiconducting superconductors \cite{Bustarret:2015}. Strontium titanate is by far the most dilute. \textbf{b)} Specific heat $C/T$, thermal conductivity $\kappa/T$, AC susceptibility $\chi_{AC}$ and resistivity $\rho$ as a function of temperature for SrTi$_{1-x}$Nb$_x$O$_3$ around the optimal doping level. The bulk $T_c$ is slightly lower than the temperature where resistivity vanishes. Abbreviation: STO, strontium titanate.}
\label{Fig3}
\end{figure*}

\subsection{The Superconducting Phase Diagram}

In the 1960s, when superconductivity was discovered in $n$-doped STO \cite{Schooley1965,Koonce1967} only a few \textit{semiconducting} superconductors were known. Since then many other superconductors whose parent is an insulator have been discovered \cite{Bustarret:2015}. The list includes celebrated insulators such as diamond, silicon and germanium. As seen in Figure \ref{Fig3}a, however, strontium titanate remains by far the most dilute among such superconductors \cite{Lin2013}. In a dilution refrigerator with a 20 mK base temperature, we have observed a transition towards zero resistivity in samples with a carrier density as low as $3\times 10^{17}$ cm$^{-3}$. By contrast, bulk superconductivity could be detected down to $4.5\times 10^{18}$ cm$^{-3}$. As seen in Figure \ref{Fig3}a, not only the superconductivity of strontium titanate is remarkably precocious, but also the dependence of its $T_c$ with carrier density does not show the steep initial increase seen in other systems. There are reports of superconductivity in Zr-substituted STO extending to even lower carrier densities \cite{Tainsh:1986,Eagles:1986}. But these are ceramic samples with dubious homogeneity. In contrast to single crystals \cite{Lin2013}, there is no evidence for quantum oscillations establishing the existence of a single continuous Fermi sea.

One puzzling feature of superconductivity in strontium titanate is that, resistive $T_c$ (the temperature at which resistivity drops to zero) and the bulk $T_c$ (the temperature at which specific heat begins to rise) are different. Figure \ref{Fig3}b shows the superconducting transition in the same optimally-doped sample monitored by four different probes. Electric resistivity vanishes well above the temperature at which specific heat, magnetic susceptibility and thermal conductivity begin to deviate from their normal-state values because of the superconducting transition. The bulk $T_c$ corresponds to the onset of the global Meissner effect seen by DC magnetization measurements \cite{collignon2017}. This means that even at optimal doping, filamentary superconductivity persists at a temperature higher than the bulk critical temperature and the same filamentary superconductivity persists in the dilute limit even after the destruction of the bulk superconductivity. One possible place where filamentary superconductivity can 'hide' is a domain wall between tetragonal domains. This speculation is supported by a recent finding reported by the Moler group who, studying Nb-doped SrTiO$_3$ thin films, found that upon warming the diamagnetic response in domain boundaries survives at a temperature $\geq 10\%$ higher than elsewhere \cite{Moler2016}. Given the well-established sensitivity of the critical temperature to strain \cite{Schooley1970} and pressure \cite{Rowley2018} (see Section \ref{QCf}), oriented strain or negative pressure in the domain boundary can be the origin of such a higher critical temperature. The polar nature of the domain boundaries \cite{Salje:1998}, in the light of the growing evidence for an intimate connection between ferroelectricity and superconductivity (see Section \ref{QCf}) may also play a role.

\begin{figure*}
\centering
\includegraphics[width=0.85\textwidth]{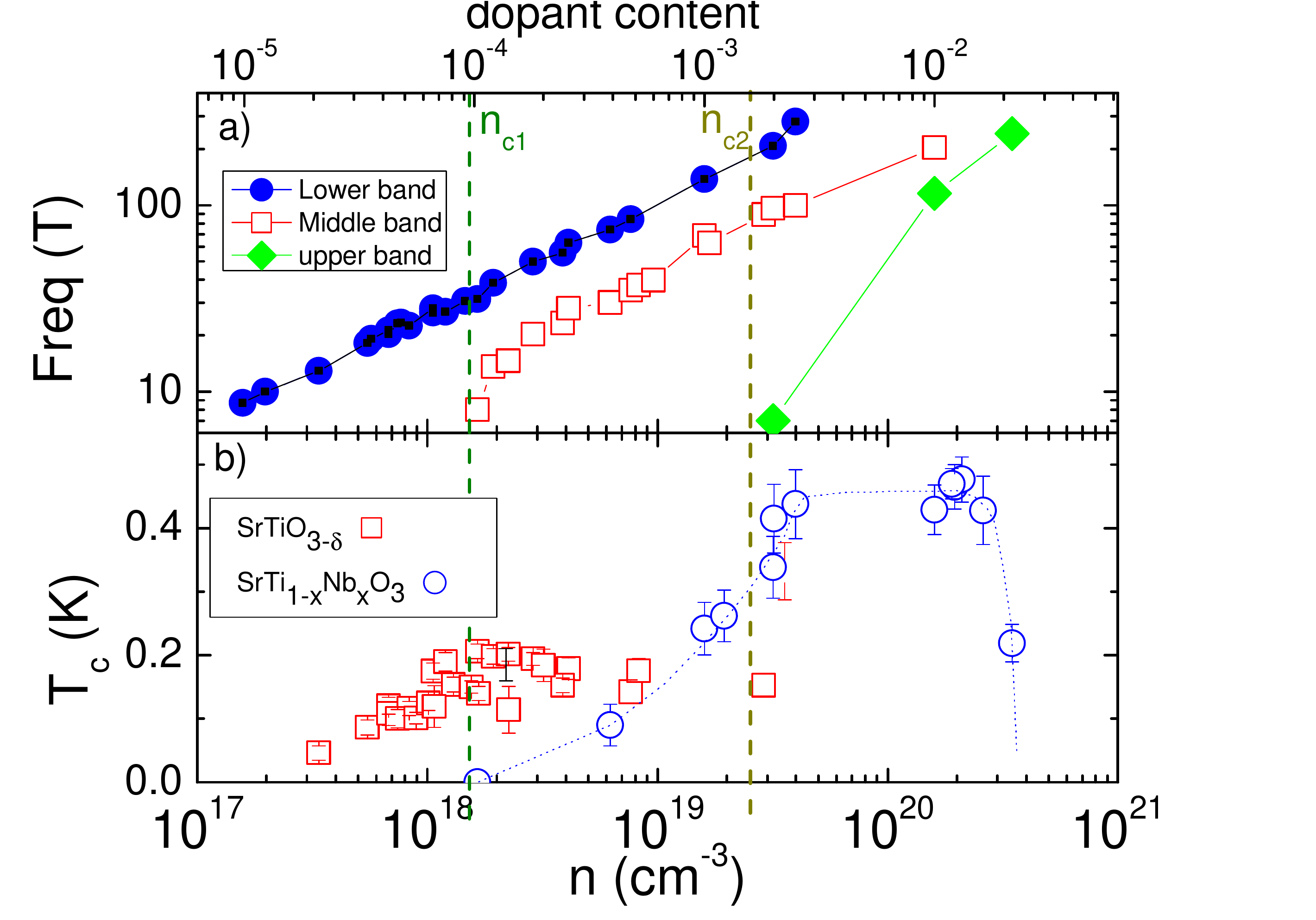}
\flushleft
\caption{\textbf{Evolution of the Fermi surface structure and resistive  $T_c$} \textbf{a)} The evolution of the frequency of quantum oscillations with carrier concentration. At two critical doping levels, $n_{c1}$ and $n_{c2}$, new frequencies emerge marking the occupation of new bands. \textbf{b)} The resistive superconducting phase diagram in SrTiO$_{3-\delta}$ and in SrTi$_{1-x}$Nb$_x$O$_3$. Note the structure, which correlates with band occupation.}
\label{Fig4}
\end{figure*}

Another puzzling feature is that dilute superconductivity, which is clearly observed in the resistivity of SrTiO$_{3-\delta}$ is absent in SrTi$_{1-x}$Nb$_x$O$_3$ (see Figure \ref{Fig4}b). The two systems do not show any detectable difference in their phonon spectrum or their Fermi surface structure. The only obvious difference is that an oxygen vacancy donates two mobile electrons and a Nb dopant only one. This may be a manifestation of the ``negative-U'' superconductivity \cite{Geballe2013}, where the presence of two electrons at the same site favors the formation of Cooper pairs in dilute SrTiO$_{3-\delta}$.

As seen in Figure \ref{Fig4}b, the superconducting dome of SrTiO$_{3-\delta}$ has a structure \cite{Lin2013,Lin2014}, that correlates with the evolution of the Fermi surface with doping (see Figure \ref{Fig4}a). The initial increase of $T_c$ in the dilute limit is interrupted as soon as the middle band begins to be occupied. Fernandes and co-authors have recently argued that the expected enhancement in $T_c$ \cite{Fernandes2013} due to the enhanced density of state by the occupation of an additional band can be suppressed in a multi-band superconductor, when the inter-band pairing interaction is repulsive \cite{Fernandes2018}.

\subsection{The Symmetry of the Order Parameter}
\label{Sec4.2}
The first study of the superconducting gap in STO was by Binnig et al. \cite{Binnig}, who performed planar tunneling experiments on SrTi$_{1-x}$Nb$_x$O$_3$ and observed two distinct gaps. Recent tunneling studies on superconducting SrTiO$_3$/LaAlO$_3$ interfaces \cite{Richter2013} and Nb-doped thin films \cite{Swartz2018}, however, could not detect such multiple gaps. These early and recent tunneling studies found that the magnitude of the gap is close to the BCS weak-field coupling value of $\Delta/k_BT_c=1.76$ and in agreement with the size of the specific heat jump \cite{Lin2014PRB}.

\begin{figure*}
\centering
\includegraphics[width=0.9\textwidth]{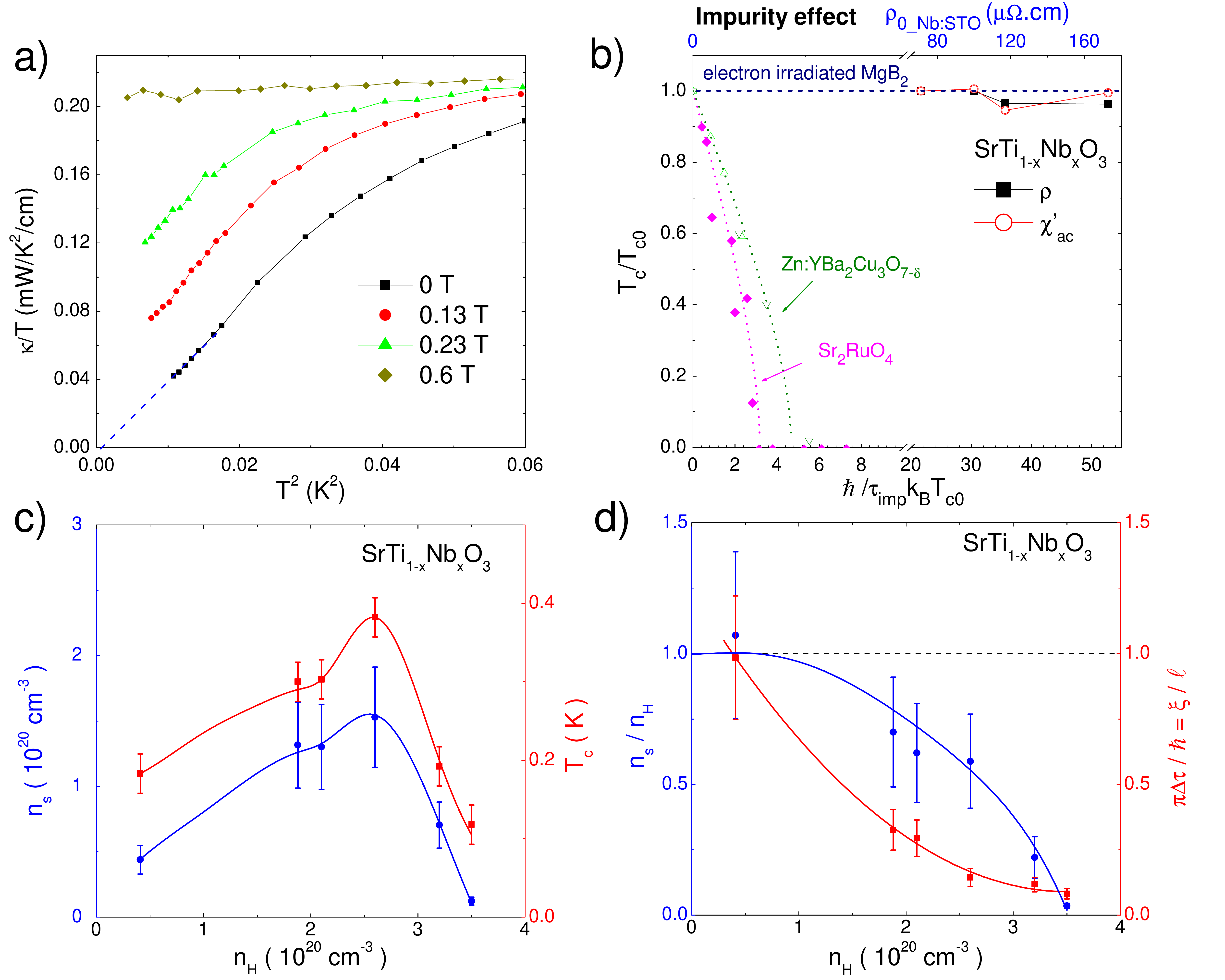}
\flushleft
\caption{\textbf{Gap structure, order parameter symmetry and superfluid density} \textbf{a)} Ratio of thermal conductivity and temperature $\kappa/T$ as a function of $T^2$. The dashed line is a linear extrapolation of the zero-field data indicating the absence of a $T$-linear term. Such a term, indicative of a residual fermionic fluid is induced by the application of magnetic field. \textbf{b)} $T_c/T_{c0}$ as a function of $\alpha=\hbar/\tau_{imp}k_B T_{c0}$, which quantifies the pair-breaking rate \cite{Millis,Abrikosov,Radtke,Tolpygo} in SrTi$_{1-x}$Nb$_x$O$_3$ compared with electron-irradiated MgB$_2$ \cite{Blinkin}, Zn-doped YBCO
 \cite{Fukuzumi} and slightly-disordered Sr$_2$RuO$_4$ \cite{Mackenzie}. The top axis shows the residual resistivity $\rho_0$ for SrTi$_{1-x}$Nb$_x$O$_3$. \textbf{c)} Zero temperature superfluid density $n_s$ and critical temperature $T_c$ as a function of the Hall carrier concentration $n_H$. \textbf{d)} The ratio of the superfluid density $n_s$ to $n_H$ and the product of the superconducting gap and the scattering time $\pi \Delta \tau / \hbar = \ell / \xi$ as a function of doping. $\Delta$ is assumed to be $1.76 k_B T_c$ and $\tau$ is extracted from low temperature resistivity \cite{collignon2017}. Abbreviation: STO, strontium titanate.}
\label{Fig5}
\end{figure*}

The study of thermal conductivity, $\kappa$, in the superconducting state \cite{Lin2014PRB} found no  $T$-linear term in $\kappa$ deep inside the superconducting state (Figure \ref{Fig5}a). Nodal quasi-particles are expected to present a  $T$-linear contribution to the heat transport \cite{shakeripour2009}. Therefore, the result implies that the superconducting gap is nodeless. Controlled disorder was found to have little effect on the superconducting transition temperature \cite{Lin2015PRB}. As seen in Figure \ref{Fig5}b, $T_c$ is insensitive to defects introduced by electron irradiation. This behavior is similar to what was observed in MgB$_2$, and contrasts with the typical sensitivity of unconventional superconductors. In d-wave YBaCuO$_{7-\delta}$ and p-wave Sr$_2$RuO$_4$, $T_c$ is extremely sensitive to potential scattering and the superconducting ground state can be completely destroyed by disorder \cite{Millis,Abrikosov,Radtke}. In contrast, according to Anderson's theorem \cite{Anderson}, Cooper pairs with s-wave symmetry are not broken by non-magnetic defects. Thus, it is safe to conclude that, at least when optimally-doped, the superconducting gap(s) is(are) s-wave and nodeless.

\subsection{Superfluid Density and Its Distribution Among Bands}

The superfluid density $n_s$, i.e., the density of carriers condensing into Cooper pairs is unusually small in this system. This quantity is linked to the penetration depth $\lambda$ by the London equation $\lambda^{-2} = \mu_0 n_s e^2/m^{\star}$ \cite{london1935}. Therefore, it can be quantified either by direct measurements of $\lambda$ or through the lower critical field $H_{c1}$ via $H_{c1} = \frac{\phi_0 }{ 4 \pi \lambda^2} ( \ln \kappa + 0.5 ) $, where $\kappa$ is the Ginzburg-Landau parameter \cite{abrikosov1957,hu1972}.

Because of its low carrier density, the expected $H_{c1}$ is of the order of the terrestrial magnetic field \cite{ambler1966,collignon2017}. Assuming that most of the superfluid density should come from the lower band and taking the effective mass extracted from quantum oscillations and specific heat (see Section \ref{Sec3.2}), it is possible to derive $n_s$ from $H_{c1}$ measurements \cite{collignon2017}. The $n_s$ dependence on doping is shown in Figure \ref{Fig5}c and follows a dome-like structure which seems to scale with $T_c$.

In a naive picture, when $T \rightarrow 0$, all carriers contribute to superconductivity and as a result $n_s$ should match the normal state carrier density $n_H$. This is indeed true in a large variety of superconductors in (or close to) the clean limit \cite{collignon_phD,collignon2017}. Nevertheless, in presence of a superconducting dome, this equality between $n_s$ and $n_H$ breaks down and there is an apparent scaling between $n_s$ and $T_c$, which can be understood thanks to the Homes law \cite{homes2005}: because of scattering, states of characteristic energy higher than $2 \Delta$ do not condense, giving the proportionality $n_s / n \propto \Delta \tau \, (\sim \ell / \xi)$ in the the dirty limit.

To take a closer look, the ratios $n_s / n_H$ and $\pi \Delta \tau / \hbar$ reported in Ref \cite{collignon2017} are plotted in Figure \ref{Fig5}d. At the lowest doping, $n_s$ matches $n_H$ as expected. Then, upon entering the dirty limit ($\ell / \xi < 1$), $n_s$ deviates from $n_H$ and becomes only a few percent of $n_H$. However, there is no strict proportionality between $n_s / n_H$ and $\Delta \tau$, as expected according to the Homes law in a single-band picture.

The latter observation may be connected to the unsettled issue of multi-band superconductivity. At optimal doping, the normal carriers are distributed among three distinct Fermi surfaces. However, recent tunneling studies detect a single superconducting gap \cite{Swartz2018} and microwave measurements \cite{thiemann2017} see a single energy scale for the superfluid condensate. A future challenge is to reconcile these experimental results with the temperature dependence of $H_{c1}$ \cite{collignon2017} in a quantitative way.

\subsection{Quantum-Critical Ferroelectricity}
\label{QCf}
\begin{figure*}
\includegraphics[width=0.9\textwidth]{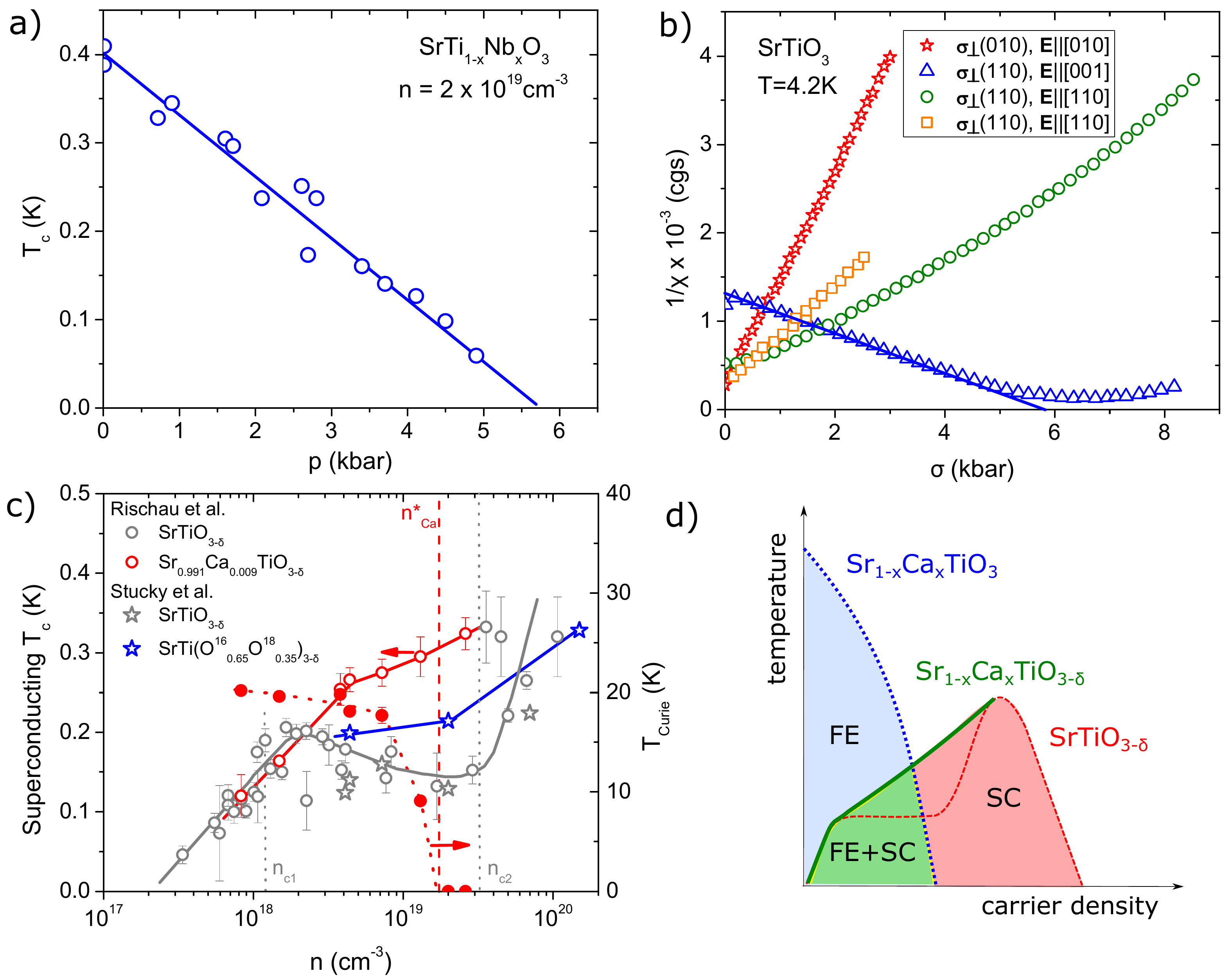}
\caption{\textbf{Superconductivity and Ferroelectricity} \textbf{a)} The dependence of critical temperature on pressure \cite{Rowley2018}. \textbf{b)} The dependence of electric permitivitty on strain \cite{Unoki:1967}. \textbf{c)} Superconducting critical temperature $T_c$ as a function of carrier concentration $n$ for SrTiO$_{3-\delta}$ \cite{Rischau:2017}, Sr$_{0.991}$Ca$_{0.009}$TiO$_{3-\delta}$ \cite{Rischau:2017} and SrTi${({}^{16} \textnormal{O}_{0.65}{}^{18} \textnormal{O}_{0.35})}_{3-\delta}$ \cite{Stucky:2016}. The Curie temperature in Sr$_{0.991}$Ca$_{0.009}$TiO$_{3-\delta}$ is also plotted. \textbf{d)} A schematic phase diagram of the different phases. Abbreviation: FE, ferroelectricity; SC, superconductivity.}
\label{Fig6}
\end{figure*}

Is it a mere coincidence that this unusual superconducting dome occurs in a doped semiconductor whose insulating parent is an incipient ferroelectric? One suspects the answer to this question to be negative. This suspicion is reinforced by the simple observation illustrated in Figure \ref{Fig6}. A small pressure of 5 kbar destroys superconductivity (panel a) and stabilizes the ferroelectric order (panel b). It is hard to see how such a small pressure could affect longitudinal phonons or their well-documented \cite{Swartz2018} strong coupling with mobile electrons. The fact that a small and comparable pressure is required to harden the soft mode and to destroy the superconductor is indicative of an intimate link between the two.

Following Rowley \emph{et al.} \cite{Rowley:2014} pointing to ferroelectric quantum criticality, Edge \emph{et al}. proposed a specific scenario in which the maximum $T_c$ is set by the emergence of the ferroelectric instability \cite{Edge:2015}. As seen in Section \ref{QP}, quantum fluctuations suppress the emergence of a long-range ferroelectric order in the pristine insulating STO. When ${}^{18}$O replaces ${}^{16}$O or a small fraction of Sr atoms is substituted by Ca atoms, a long-range order appears. Furthermore, the composition of 35\% ${}^{18}$O has been identified as a quantum critical point \cite{Rowley:2014}. According to the quantum critical scenario, the maximum $T_c$ enhances and the dome shifts to lower doping levels upon ${}^{18}$O substitution, as isotopic substitution pushes the quantum critical point into the electron doping region that allows superconductivity \cite{Edge:2015}.

Experimental support for the ferroelectric quantum-critical scenario, the subject of several recent theoretical studies \cite{Rowley2018,Arce2018,Kansugi2018,Wolfle2018}, has recently emerged in the studies of SrTi${({}^{16} \textnormal{O}_{0.65}{}^{18} \textnormal{O}_{0.35})}_{3-\delta}$ \cite{Stucky:2016} and Sr$_{0.991}$Ca$_{0.009}$TiO$_{3-\delta}$ \cite{Rischau:2017}. As seen in Figure \ref{Fig6}c, both Ca \cite{Rischau:2017} and ${}^{18}$O \cite{Stucky:2016} substitution were found to enhance the superconducting critical temperature in a finite window of doping.

Sr$_{1-x}$Ca$_{x}$TiO$_{3-\delta}$ ($0.002<x<0.02$) is a metal, and therefore not a true a ferroelectric with a reversible bulk polarization. Nevertheless, it hosts a phase transition structurally indistinguishable from the ferroelectric transition occurring in the insulator without oxygen vacancies \cite{Rischau:2017}. Figure \ref{Fig6}c plots the Curie temperature detected in these metallic samples showing that there is a region in the phase diagram in which superconductivity and ferroelectricity coexist (sketched in Figure \ref{Fig6}d). The enhancement of $T_c$ is the strongest in the vicinity of the destruction of the ferroelectric-like order, in qualitative agreement with the quantum critical scenario. Hence, a major role played by the soft TO phonon mode in the formation of Cooper pairs near the peak of the superconducting dome is confirmed by these experiments. The survival of zero-resistivity (accompanied or not by bulk superconductivity) in oxygen-reduced strontium titanate remains a challenge to theory. A promising possible candidate are plasmons \cite{Takada:1980,Ruhman2016}.

\section{Concluding Remarks}
The BCS theory based the phenomenon of superconductivity on solid theoretical grounds. But it also led to a paradigm shift. In the words of A. J. Leggett, superconductivity "was understood as occurring almost inevitably in a very degenerate Fermi system" whenever phonon-mediated attraction outweighs Coulomb repulsion \cite{Leggett2011}. Superconductivity of strontium titanate constitutes an intriguing case of the survival of a superconductor in an unexpected context where carriers have a small kinetic (and even smaller potential) energy. This strangeness highlights the limits of \textit{ab initio} theory in determining the outcome of the balance between attraction and repulsion and the irony of a glorious scientific theory weak in prediction and strong in postdiction, as underlined by the recent discovery of superconductivity in bismuth \cite{Prakash2017}.
\section*{DISCLOSURE STATEMENT}
The authors are not aware of any affiliations, memberships, funding, or financial holdings that might be perceived as affecting the objectivity of this review.

\section*{ACKNOWLEDGMENTS}
We thank A. Balatsky, P. Chandra, D. Eagles, J. Flouquet, J. Hemberger, G. G. Lonzarich, T. Lorenz, D. Maslov and N. Spaldin and D. van der Marel for discussions and comments. This work was supported by Fonds-ESPCI and by the Agence Nationale de Recherche as a part of the QUANTUMLIMIT project. BF acknowledges support from Jeunes Equipes de l'Institut de Physique du Coll\`ege de France (JEIP).

\section*{SUMMARY POINTS}
\begin{enumerate}
\item SrTiO$_3$ is a quantum paraelectric with a very large dielectric permittivity. The ferroelectric order can be stabilized by applying strain, chemical (Ca for Sr) or isotopic (O$^{18}$ for O$^{16}$) substitution.
\item The large permittivity of the insulator leads to a long effective Bohr radius for dopants, which can easily turn the system into a dilute metal. Charge transport in this metal presents intriguing features including the persistence of  $T$-square resistivity in the extreme dilute limit and a non-saturating magnetoresistance at room-temperature and above.
\item The superconducting dome persists down to very low carrier densities where several usual assumptions in the BCS scenario cease to be valid.
\item Superconductivity and ferroelectricity are intimately linked and there is experimental support for the quantum-critical scenario according to which the critical temperature is boosted by the vicinity to the ferroelectric order.
\end{enumerate}

\section*{FUTURE ISSUES}
\begin{enumerate}
\item A reliable theory of charge transport in the dilute metallic state is missing.
\item Do the electrons become heavier with warming or do they keep their low-temperature mass?
\item Can a quantum-critical scenario explain the persistence of superconductivity in the very dilute limit in the oxygen-reduced strontium titanate?
\item Where does the difference between the bulk and resistive  $T_c$ come from?
\item Is the system a multi-gap superconductor?
\end{enumerate}


\bibliography{bib} 
\end{document}